\documentclass[reprint,aip,jcp,preprintnumbers,superscriptaddress,twocolumn,floatfix]{revtex4-1}

\usepackage[utf8]{inputenc}
\usepackage[T1]{fontenc}
\usepackage{graphicx}
\usepackage{color}
\usepackage{bm}
\usepackage{amsmath}
\usepackage{xspace}
\usepackage{bbold}
\usepackage{units}
\usepackage{dcolumn}
\usepackage{paralist}
\usepackage{natmove}
\usepackage{amssymb}
\usepackage{setspace}
\usepackage{mathtools}
\usepackage{gensymb}
\usepackage{upgreek}

\DeclareMathAlphabet{\mathdutchcal}{U}{dutchcal}{m}{n}
\SetMathAlphabet{\mathdutchcal}{bold}{U}{dutchcal}{b}{n}
\DeclareMathAlphabet{\mathdutchbcal}{U}{dutchcal}{b}{n}

\usepackage{tikz}
\usepackage{xcolor}

\begin{document}
\title{Free-energy landscape of polymer--crystal polymorphism}
\author{Chan Liu}
\affiliation{Max Planck Institute for Polymer Research, 55128 Mainz,
Germany}
\author{Jan Gerit Brandenburg}
\affiliation{Interdisciplinary Center for Scientific Computing,
University of Heidelberg, 69120 Heidelberg, Germany}
\affiliation{Digital Organization, Merck KGaA, 64293 Darmstadt,
Germany}
\author{Omar Valsson}
\affiliation{Max Planck Institute for Polymer Research, 55128 Mainz, 
Germany}
\author{Kurt Kremer}
\affiliation{Max Planck Institute for Polymer Research, 55128 Mainz, 
Germany}
\author{Tristan Bereau} 
\email{t.bereau@uva.nl}
\affiliation{Max Planck Institute for Polymer Research, 55128 Mainz, 
Germany}
\affiliation{Van 't Hoff Institute for Molecular Sciences and
Informatics Institute, University of Amsterdam, Amsterdam 1098 XH, The
Netherlands}

\date{\today}
\begin{abstract}
Polymorphism rationalizes how processing can control the final
structure of a material. The rugged free-energy landscape and
exceedingly slow kinetics in the solid state have so far hampered
computational investigations. We report for the first time the
free-energy landscape of a polymorphic crystalline polymer,
syndiotactic polystyrene. Coarse-grained metadynamics simulations
allow us to efficiently sample the landscape at large. The free-energy
difference between the two main polymorphs, $\alpha$ and $\beta$, is
further investigated by quantum-chemical calculations. The two methods
are in line with experimental observations: they predict $\beta$ as
the more stable polymorph at standard conditions. Critically, the
free-energy landscape suggests how the $\alpha$ polymorph may lead to
experimentally observed kinetic traps. The combination of multiscale
modeling, enhanced sampling, and quantum-chemical calculations offers
an appealing strategy to uncover complex free-energy landscapes with
polymorphic behavior.
\end{abstract}

\maketitle

\section{Introduction}

The complex interplay of molecular interactions can lead to a material
exhibiting multiple distinct forms in its solid
state~\cite{brittain1999polymorphism}. This
polymorphism often results in widely different materials properties,
making the study of polymorphism both essential for quality control in
manufacture, but also a fascinating structure--property problem. Beyond
structure and property, the intermediate processing of the material
has a key impact on the resulting polymorph. Fundamentally this stems
from two ingredients: ($i$) the underlying free-energy landscape is
sufficiently rugged to display several low-lying metastable states;
and ($ii$) exceedingly slow kinetics exhibited in the solid phase,
preventing a full/ergodic kinetic relaxation.

The screening of polymorphs has traditionally exclusively been
performed experimentally, in spite of the significant costs involved.
Computational methods hold the promise of predicting polymorphic
stability before going to the laboratory. In the context of molecular
crystals, especially targeted at pharmaceuticals and porous (organic)
cages, a considerable body of work has recently
emerged~\cite{gdanitz1998ab, bernstein2002polymorphism, Jansen2010,
Jones2011, Abramov2012, PyzerKnapp2014, CruzCabeza2015, Neumann2015,
Price2016, price2017molecular, Pulido2017, Day2017}. The modeling of
polymorphism holds two challenges: sampling and modeling accuracy. The
free-energy landscape exhibits an overwhelming number of
configurations, of which only an infinitesimal fraction competes in
terms of low-lying states. Furthermore, correctly ranking the relative
free energies of each conformer requires computational methods that
are accurate enough (of the order of thermal energy, $k_{\rm B}T$) to
reproduce the underlying interactions. Many methods exist to tackle
these two challenges, out of which we mention the use of an
appropriately-tuned force-field based method for the sampling and
electronic-structure methods (e.g., density functional theory) for the
energetic characterization~\cite{reilly2016report, Marom2013,
hoja2019reliable}.

Here, we focus on polymers, which not only embody countless industrial
applications, but also for which we critically lack a detailed picture
of its free-energy landscape. Evidently, the increased number of atoms
per molecule involved will lead to higher structural correlations,
significantly larger barriers, and much longer timescales compared to
molecular crystals of small molecules. While many semicrystalline
polymers possess only a single type of unit cell, there are
exceptions: syndiotactic polystyrene (sPS) for instance is well-known
for its complex crystal polymorphism.

\begin{figure}[htbp]
    \centering
    \includegraphics[width=0.8\linewidth]{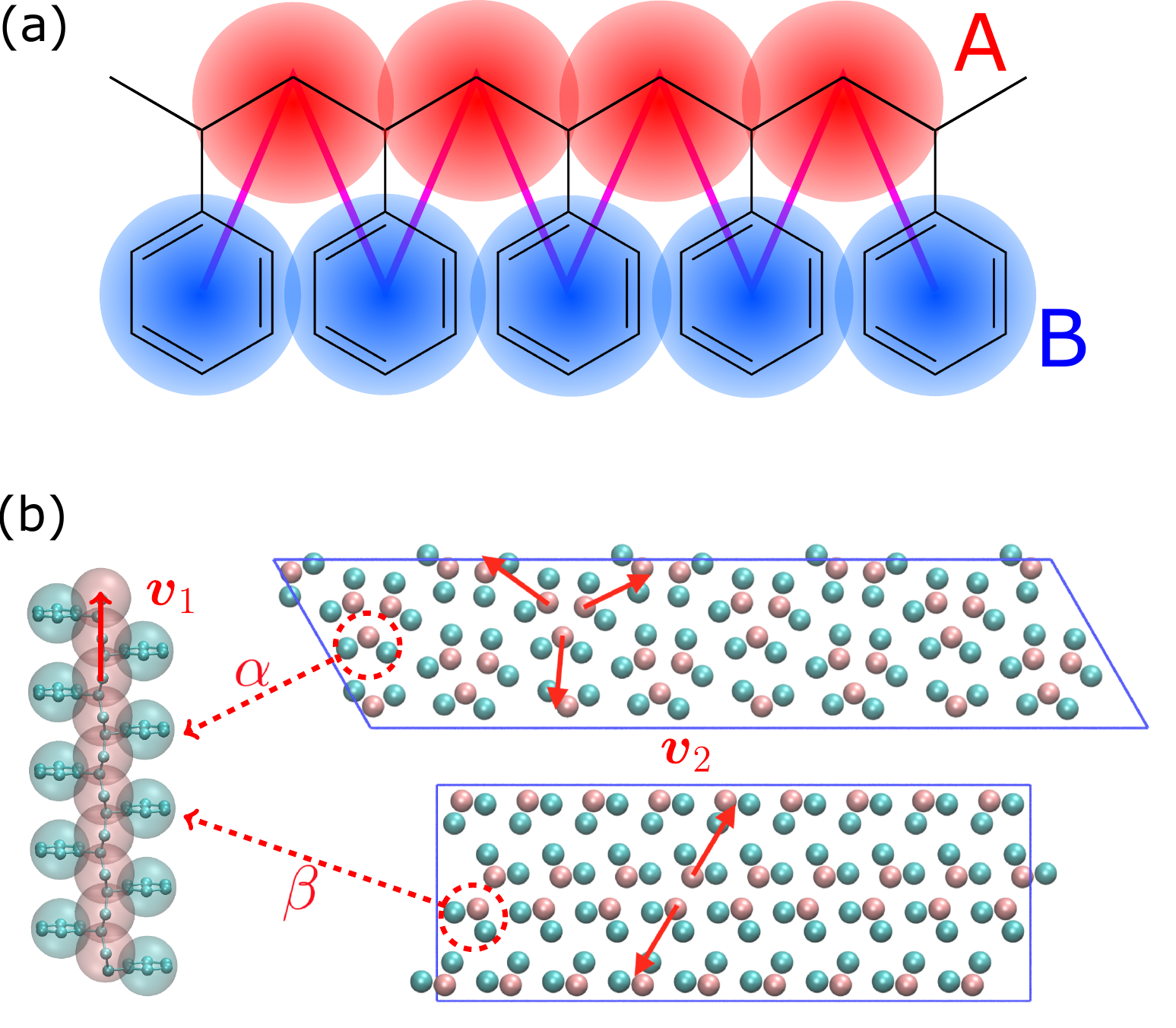}
    \caption{(a) Molecular structure of polystyrene and CG-mapping
    scheme of the Fritz model~\cite{Fritz2009}. (b) Left: Longitudinal
    view of an all-trans chain conformation; Right: transverse section
    of the experimentally-resolved $\alpha$ and $\beta$ forms. In the
    transverse sections, each polymer chain is represented by three CG
    beads (highlighted in red dashed circles). The red solid lines
    represent cross-sectional vectors pointing from the backbone to
    the bisector of its two closest side chains.} 
\label{fig:mol-vec}
\end{figure} 

In this work, we aim at unraveling the underlying free-energy
landscape of sPS, to better understand the body of experimental
evidence gathered around its polymorphs. We focus on the
thermally-induced processing aspect---the $\alpha$ and $\beta$
polymorphs, shown in Fig.~\ref{fig:mol-vec}~\cite{Rosa1991,
Corradini1994, Rosa1996, Cartier1998, Rosa1992, Chatani1993}.  They
share the same intrachain conformations, but with different
interchain-packing structures. The $\alpha$ and $\beta$ forms of sPS
are further classified into limiting disordered forms, $\alpha'$ and
$\beta'$, and limiting ordered forms, $\alpha''$ and
$\beta''$~\cite{Guerra1990, Rosa1991, Rosa1992, Rosa1996}. The
processing conditions impact the forms experimentally
observed~\cite{Guerra1990, Rosa1992}. Fig.~\ref{fig:mol-vec} displays
the limiting ordered forms. 
%

Experimental evidence has pointed out the strong impact of the
processing conditions: starting material, initial temperature, and
cooling rate~\cite{Guerra1990, Rosa1991, Rosa1992, Sun1999, Woo1999,
Lin2000, Sun2001}. They hint at a highly complex free-energy
landscape. The relative stability between the $\alpha$ and $\beta$
polymorphs is noteworthy: ($i$) Starting from a high initial
temperature, a fast (slow) cooling will lead to the $\alpha$ ($\beta$)
forms; ($ii$) Under identical slow cooling rate, melt crystallization
starting under 230\celsius\xspace and above 260\celsius\xspace will
yield the $\alpha$ and $\beta$ polymorphs, respectively, while
intermediate temperatures generate mixtures thereof. These results
suggest that given sufficient mobility thanks to a high initial
temperature and a slow enough cooling rate, the preferred packing
structure corresponds to the $\beta$ form. In case of stiffened chains
and/or reduced molecular mobility, the $\alpha$ polymorph is
preferred. These glimpses at the structural properties of sPS indicate
that crystallization to the $\alpha$ form results from a
kinetically-controlled process, while $\beta$ would be the
thermodynamically stable form. Our understanding thereby falls short
in several ways: how the free-energy landscape translates into the
apparent differences between the two forms, but also a more
mechanistic insight as to the origin of these effects. 

While computational studies of polymer crystal polymorphs remain to
date extremely limited, we note the work of Tamai and co-workers on
nanoporous cavity structures in the $\alpha$ and $\beta$
forms~\cite{Tamai2003}, the diffusion of gases~\cite{Milano2002,
Tamai2003-1}, the orientational motion of guest
solvents~\cite{Tamai2003-2, Tamai2005}, and the adsorption of small
molecules~\cite{Figueroa2010, Figueroa2010-1, Sanguigno2011}. These
studies helped understand structural features of some of these forms.
Unfortunately the atomistic resolution involved strongly limit the
timescale that can be reached with the simulations---on the order of
nanoseconds. This prevents both the observation of self assembly, but
also polymorph interconversion, thereby hindering access to the
free-energy landscape.

To address the time-scale issue, we turn to coarse-grained (CG)
modeling. By lumping several atoms into one larger superparticle or
bead, CG models can sample significantly faster, while offering a
systematic connection to the reference chemistry~\cite{Noid2013-1}.
Some of us recently applied a structure-based CG model aimed at
reproducing certain thermodynamic aspects of sPS~\cite{Fritz2009}.
Despite a parametrization and validation performed exclusively in the
melt, we found remarkable transferability to the crystalline phase:
not only does the CG model stabilize the $\alpha$ and $\beta$
polymorphs, the melting temperatures of the two phases were found to
be in excellent agreement~\cite{Liu2018}. Our study aimed at an
exploration of the self-assembly mechanisms of sPS, using a
temperature-based enhanced-sampling molecular dynamics (MD)
techniques---parallel tempering. In the present work, we instead turn
to methods based on collective variables (CVs), specifically
metadynamics~\cite{Laio2008, valsson2016enhancing}. We will show that
an appropriate choice of CVs can lead to convergence of the
simulations, and offer us access to a great diversity of structural
forms. We present for the first time the free-energy landscape of
polymorphism of a polymer crystal.

We further challenge the calculations of the free-energy difference
between $\alpha$ and $\beta$ polymorph stability by means of
quantum-chemical calculation at the density functional theory (DFT)
level. The results show excellent agreement with the CG simulations
given the change in resolution. Critically, we find in both cases the
preferential stabilization of the $\beta$ phase---in line with
experiments.

\section{Results and Discussion}

\subsection{Metadynamics}

In the Methods we present collective variables (CVs) that are capable
of distinguishing five different phases of sPS (Sec.~\ref{sec:CV}). A
significant distinction between these phases is essential to also
enable the discovery of other intermediate phases. A further
requirement is the absence of hidden barriers that would hinder
dynamics along the CVs~\cite{Laio2008}. 
To alleviate possible artifacts due to an impropriate choice of CVs,
we test several of them and later reweight all simulations to the same
CV space. This further allows us to empirically check the convergence
of our simulations.

We focus on a two-dimensional CV exploration, as a balance between
exploration and convergence: a three-dimensional CV-space exploration
can require excessive memory and presents challenges to converge due
to the curse of dimensionality. We note that extensions of the method,
such as bias-exchange and parallel-bias metadynamics, can help along
these lines~\cite{Piana2007, Marinelli2009, Baftizadeh2012,
Pfaendtner2015}. We herein present four combinations of CVs referenced
in Tab.~S1: 
($i$) $\Delta\mathcal{S}$ \& $\mathcal{S}_1$; ($ii$)
$\Delta\mathcal{S}$ \& $\mathcal{S}_2$; ($iii$) $\Delta\mathcal{S}$ \&
$\mathcal{S}_3$; ($iv$) $\Delta\mathcal{S}$ \& $P_2({\bm v}_2)$.

Fig.~S5 shows a metadynamics simulation at $T=400$\,K driven by the
combination of CVs $\Delta\mathcal{S}$ \& $P_2({\bm v}_2)$ using two
walkers. The walkers were initiated from the two crystalline phases
$\alpha$ and $\beta$.  The simulations are able to transition many
times between the main polymorphs, starting at around 20\,ns
(Fig.~S5a). 
The results highlight that the ability of the CVs to distinguish
$\alpha$ from $\beta$ help ensure large conformational transitions
(Fig.~S2). 
We note however the absence of conformations in the amorphous phase,
due both to our sampling below the transition temperature (roughly
450\,K) and our protocol's restraining box range and chain direction
(see Sec.~\ref{sec:metad}).

\subsection{Polymorphic stability}

\begin{figure}
    \centering
    \includegraphics[width=0.75\linewidth]{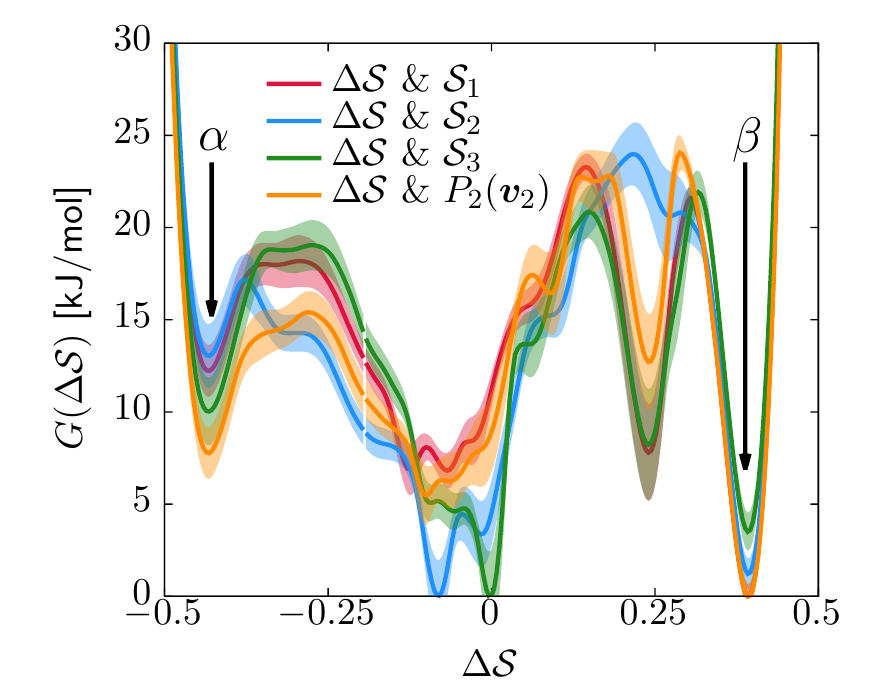}
    \caption{Convergence of the free-energy calculations. Comparison
    of four metadynamics simulations at $T=400\,$K with different CVs
    (see labels) projected on $\Delta\mathcal{S}$.}
    \label{fig:fes-error}
\end{figure}

\begin{figure}
    \centering
    \includegraphics[width=0.85\linewidth]{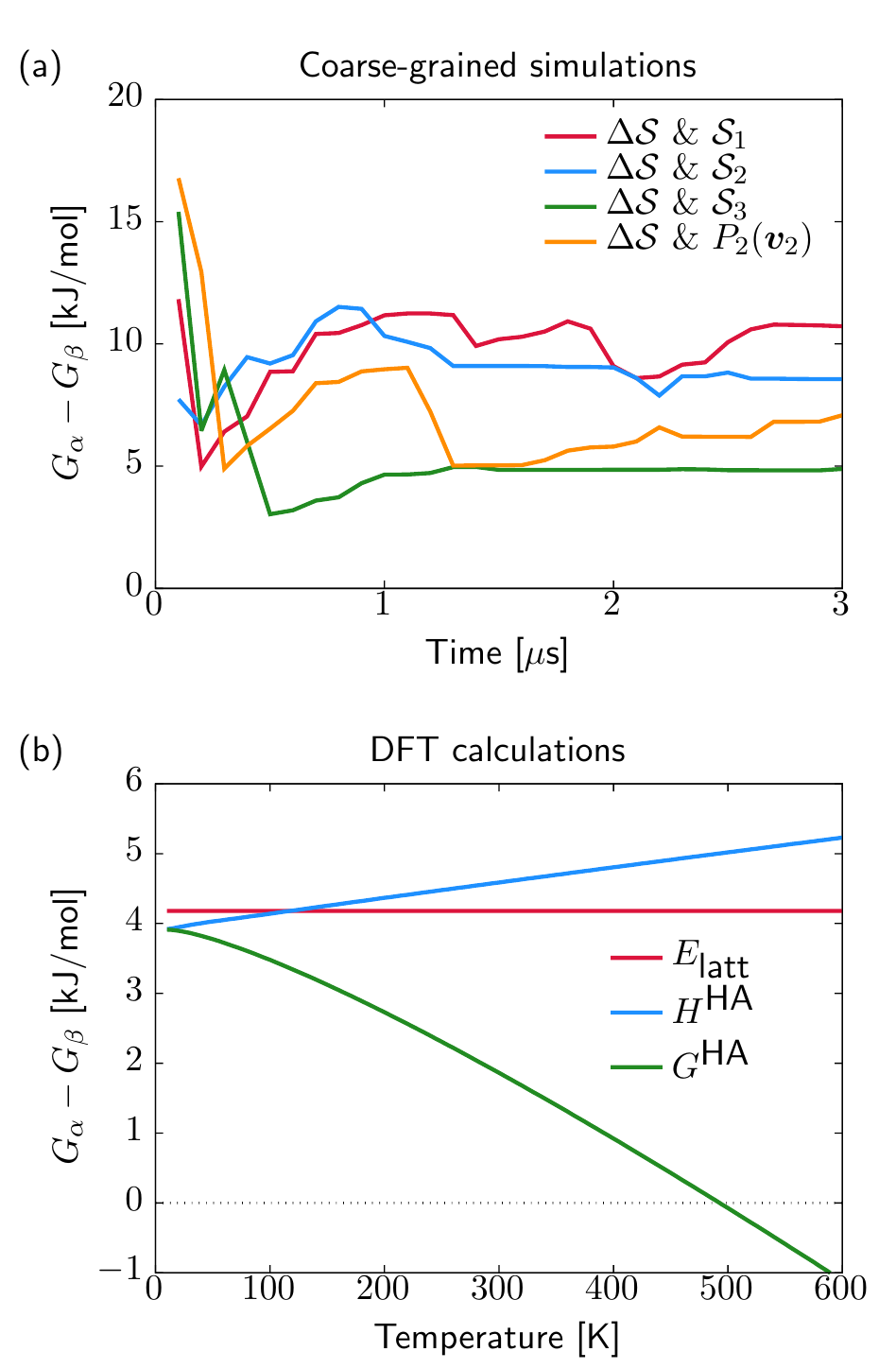}
    \caption{Free-energy difference between $\alpha$ and $\beta$
    forms. (a) The CG simulations show the time evolution of $G_\alpha
    - G_\beta$ from the different Metadynamics simulations. (b) The
    DFT-based stability differences at standard pressure and as a
    function of temperature, including lattice energy
    $E_\textup{latt}$, enthalpy $H^\textup{HA}$, and free energy
    $G^\textup{HA}$.}
    \label{fig:comp_cg-dft}
\end{figure}

Having established that our combinations of CVs enables a satisfactory
transition between major polymorphs, we turn to the question of
convergence. To assess convergence, we run metadynamics in the CV
space of our four combinations, and marginalize the free-energy
landscape to only display their common CV: the SMAC difference
$\Delta\mathcal{S}$. Fig.~\ref{fig:fes-error} compares the free-energy
surface as a function of the CV, $G(\Delta\mathcal{S})$. We have
marked the $\alpha$ and $\beta$ polymorphs according to values of the
CV from unbiased MD simulations (see SI). All four curves agree within
roughly 5~kJ/mol across the range of $\Delta\mathcal{S}$ values,
despite their sampling along different complementary CV. Convergence
as a function of simulation time is further displayed in
Fig.~\ref{fig:comp_cg-dft}a, which focuses on the free-energy
difference between the $\alpha$ and $\beta$ polymorphs, $G_\alpha -
G_\beta$. We find that all curves converge after roughly 1 to
2~$\mu$s. We do see variations between simulations reminiscent of the
spread in panel (a). Given the remarkable complexity of probing the
free-energy landscape of polymer crystals, we consider this level of
agreement encouraging indicators of the level of convergence of our
simulations.

Metadynamics simulations on different system sizes lead to similar
free-energy profiles (see Fig.~S6), whether changing the number of
chains in the box or the number of monomers per chain. We rationalize
the lack of system-size dependence in two ways: ($i$) The lack of
dependence in the number of chains can be explained by the collective
nature of the transitions, where the box is so small that all chains
transition to a new phase at once; and ($ii$) The free-energy barriers
do not scale with the number of monomers, because the transitions are
orthogonal to the chain director, as indicated by the order parameters
$P_2({\bm v}_1)$ and $P_2({\bm v}_2)$ (see Fig.~S2).

Having identified the two polymorphs as local minima with a
rationalization of kinetic routes in between them, we try to further
establish their relative thermodynamic stability. For this, we use the
structures identified in the dynamics and employ quantum mechanical
methods for local structure relaxations and prediction of their
temperature dependent free energy (see section~\ref{sec:dft}). In
contrast to the molecular dynamics simulations, our DFT results do not
describe anharmonicities of the energy surface, potentially neglecting
some thermal effects. On the other hand, the described interactions
are at a quantum mechanical level, physically more sound, and thus
expected to be more accurate compared to the classical potentials used
in our molecular dynamics. We show in Fig.~\ref{fig:comp_cg-dft}b the
quantum-mechanical energy difference between the two polymorphic
forms. In a static picture at 0\,K, sPS $\beta$ is predicted to be
more stable by 4\,kJ/mol. Heating to 400\,K, the enthalpy difference
increases by 1\,kJ/mol, while the free energy difference decreases by
3 kJ/mol. At elevated temperatures of about 500\,K, form $\alpha$ is
predicted to be more stable, which is, however, not experimentally
observable due to the amorphous phase. The stability difference at
room temperature is in excellent agreement with the estimations from
our dynamics simulations indicating a stabilization of $\beta$ by 5-10
kJ/mol. The smaller stability difference predicted by DFT is in line
with our experience from molecular crystals, where higher quality
interaction energy models typically lead to smaller energy gaps
between polymorphs.\cite{reilly2016report} The overall analysis
matches the experimental expectation that form $\alpha$ crystallizes
by rapid cooling from elevated temperatures, while $\beta$ forms in
slow cooling experiments~\cite{Woo2001}.

The analysis reported in Fig.~\ref{fig:comp_cg-dft} shows that the
$\beta$ form is systematically better stabilized than the competing
$\alpha$ form. The 1D landscape clearly separates $\alpha$ from
$\beta$ at the left and right sides of the range, respectively. These
are separated by both $\alpha/\beta$ mixtures and the amorphous phases
at around $\Delta\mathcal{S} \approx 0$. Interestingly, we observe a
significantly lower free-energy barrier upon going from the amorphous
phase to the pure $\alpha$ polymorph, than the $\beta$ polymorph:
while the former is between 10 and 15~kJ/mol high, the other is
upwards of 20~kJ/mol. The $\beta$ form is more stable across the CV
space, but the $\alpha$ form is easier to \emph{reach} from the
mixture and amorphous phases---a kinetic effect. This can help
rationalize the kinetic-trap behavior of the $\alpha$ form found
experimentally~\cite{Woo2001}. Some of us had previously identified an
overpopulation of the $\alpha$ form when probed in simulation box
geometries concomitant to the $\alpha$ unit cell, suggesting a
templating mechanism~\cite{Liu2018}.

\subsection{Free-energy landscape of sPS}

\begin{figure*}
    \centering
    \includegraphics[width=0.9\linewidth]{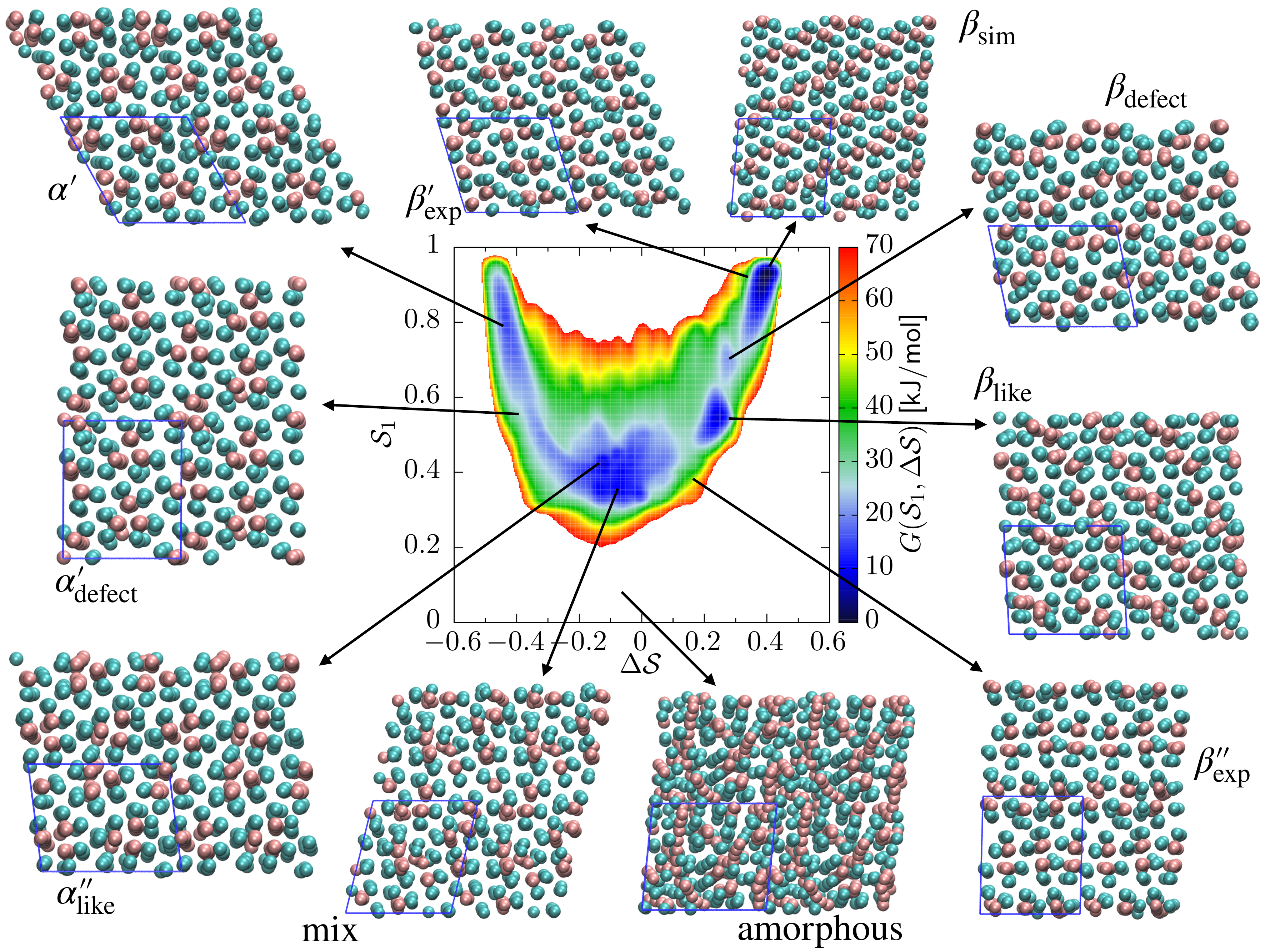}
    \caption{Free-energy landscape of sPS sampled with metadynamics as
    a function of the CVs: $G(\mathcal{S}_1, \Delta\mathcal{S})$.
    Various structures---representative of the simulations and/or the
    experiments---are displayed and identified on the surface.}
    \label{fig:fes-2d}
\end{figure*}

As an extension to Fig.~\ref{fig:fes-error}, Fig.~\ref{fig:fes-2d}
shows a representation of the free-energy landscape for the CV
combination $\Delta\mathcal{S}$ \& $\mathcal{S}_1$. Stability is color
coded from blue to red. We observe a large diversity of phases with
distinct structural features. Notably, we find many more structures
than our previous study based on parallel tempering~\cite{Liu2018}.

We first analyze structures similar to the $\alpha$ polymorph. Of
particular interest are symmetries around triplets of chains that form
the fundamental symmetric unit of $\alpha$ structures (see
Fig.~\ref{fig:mol-vec}b). As compared to our previous work that used
parallel tempering, we observe a broader variety of relative
orientations between chain triplets: Small but noticeable variations
can be found among the $\alpha$-type structures on the landscape. The
differentiation between $\alpha'$ and $\alpha''$ is made more
difficult by imposing 12 chains in our simulation box, while the unit
cell of $\alpha$ contains only 9 chains~\cite{Rosa1996}. We emphasize
the difference between apparently distinct forms stabilized in our
simulations: $\alpha'$, $\alpha'_\mathrm{defect}$ and
$\alpha''_\mathrm{like}$, shown in Fig.~\ref{fig:fes-2d}. All triplets
of chains, represented by groups of tan-colored beads, exhibit
virtually identical orientations in $\alpha'$ and little variation in
$\alpha'_\mathrm{defect}$: the angles between side-chain vectors are
almost strictly at 120$\degree$, leading to $\Delta\mathcal{S} \approx
-0.4$. $\alpha''_\mathrm{like}$, on the other hand, displays two
different triplet orientations, analogous to the experimentally
resolved $\alpha''$. The angles between side-chain vectors does not
always correspond to 120$\degree$, leading to $\Delta\mathcal{S}
\approx -0.2$, located near the mixture phases. Our simulations do
not stabilize the $\alpha''$ polymorph, which may be due to the number
of chains incongruent with the unit cell, or possibly limitations in
the CG force field in reproducing fine steric
features~\cite{Fritz2009, Liu2018}. 

In line with our previous study, we find an alternate form to the
experimentally-resolved~\cite{Guerra1990,Rosa1992} limiting disordered
$\beta'$ and limiting ordered $\beta''$ polymorphs as the most global
minimum of sPS: $\beta_{\rm sim}$, where we highlighted the difference
in layering~\cite{Liu2018}. While the parallel tempering simulations
led to neither experimental form, the metadynamics simulations
successfully sampled them, albeit at too high free energy: the
$\beta'_{\rm exp}$ and $\beta''_{\rm exp}$ forms sit at about 30 and
50~kJ/mol higher than the global minimum, respectively. We argued
before that the simple description of the side-chain sterics likely
had a detrimental effect on the stability of the $\beta$ polymorphs.
This effect was motivated by a discrepancy in the melting temperature
from CG simulations, as compared to reference atomistic simulations:
in excellent agreement for the $\alpha$ form, but too-low stability
for $\beta$. In the context of the present work, these structural
artifacts likely give rise to shifts in the free-energy landscape
shown in Fig.~\ref{fig:fes-2d}.


\section{Conclusion}

In this paper, we study the free-energy landscape of syndiotactic
polystyrene (sPS) using computational methods to get microscopic
insight into polymorphic interconversion. The development of adequate
collective variables (CVs) applied to metadynamics, together with the
use of a remarkably transferable coarse-grained (CG) model allow us
for the first time to cross the significant barriers between
polymorphs in polymer crystals. Minute structural differences between
polymorphs requires finely-tuned CVs to account for the small
variations.  Rather than relying on a single CV combination, running
metadynamics on several such combinations and reweighing them helps us
test for convergence of the simulations. We find excellent agreement
between four such combinations, despite the significant barriers
exerted by the system. 

We rely on a combination of two different SMAC
variables~\cite{Giberti2015} to build a free-energy landscape. The
$\beta$ form clearly stands as the global minimum, even though we
observe fine differences between the simulated and experimental
layerings, arguably an artifact of the side-chain representation in
the CG model. Encouragingly, we do observe the two experimental
$\beta'_{\rm exp}$ and $\beta''_{\rm exp}$ structures in the
metadynamics runs. The $\alpha$ form is between 5 and 10 kJ/mol less
stable than the global minimum. As a complementary approach, we used
quantum mechanical models to compute the temperature dependent
relative stability of the $\alpha$ and $\beta$ polymorphs. We predict
the stability to be slightly smaller (1\,kJ/mol), but can overall
confirm the CG simulations.

Remarkably, we find a significantly lower free-energy \emph{barrier}
upon going from the amorphous phase to $\alpha$ rather than $\beta$.
This lowered activation could partially explain the $\alpha$
polymorph's tendency to be identified as a kinetic trap. It also
complements our previous observation: $\alpha$ is overstabilized in
box geometries congruent with its unit cell, a templating mechanism of
sorts. Differences in nucleation rates may further help drive the
system in the direction of $\alpha$, although this is beyond the scope
of this work.

Varying the system size will naturally affect the results: smaller
simulation boxes are more likely to suffer from incompatibilities with
different crystal units, resulting in artificially low stabilization.
On the other hand, the significant free-energy barriers will become
more challenging to cross as simulation boxes grow. The
present study demonstrates that a multiscale approach can provide
insight and complementary information to experiments on
polymer--crystal polymorphism.


\section{Simulation Methods}

\subsection{Coarse-grained simulations}

We rely on a previously-developed coarse-grained (CG) model for
syndiotactic polystyrene (sPS), referred hereafter as the Fritz
model~\cite{Fritz2009}. It maps each monomer onto two types of CG
beads: ``A'' for the chain backbone and ``B'' for the phenyl ring
(Fig.~\ref{fig:mol-vec}a). The model represents PS by a linear chain
of alternating A and B CG beads, supplemented by sophisticated bonded
potentials to ensure accurate structures, including the correct
tacticity---enabling the crystallization of sPS. The bonded
interactions were obtained by direct Boltzmann inversion of
distributions obtained from atomistic simulations of single chains in
vacuum. The nonbonded potentials are derived by the conditional
reversible work (CRW) method~\cite{Brini2011,Deichmann2017}:
constrained-dynamics runs with the all-atom model of two short chains
in vacuum.

\subsection{Collective variables} 
\label{sec:CV}

Metadynamics acts on the selected CVs to drive the system and cross free-energy
barriers. This puts an essential role on the choice of CVs, known to be
extremely system and process dependent~\cite{Laio2008}. A large variety of CVs
have been used, for instance, distances, angles, or dihedrals formed by atoms or
groups of atoms~\cite{Gervasio2005}, coordination
numbers~\cite{Bussi2006,Fiorin2006,Piana2007}, and Steinhardt
parameters~\cite{Trudu2006,Quigley2008,Quigley2008-1}. 

Crystallization is typically characterized by long-range order. For
example, cluster symmetries are often probed via the Steinhardt
parameter, $Q_l$~\cite{Steinhardt1981,Steinhardt1983}, which
represents the rotationally-invariant spherical harmonics of order
$l$. $Q_4$ and $Q_6$, in particular, have been studied in the context
of Lennard-Jones particles~\cite{Trudu2006}, ice~\cite{Quigley2008},
and calcium carbonate nanoparticles~\cite{Quigley2008-1}. To study
polymorphism in sPS, however, we have found the Steinhardt parameter
to inefficiently distinguish crystalline forms (see Fig. S3). We
rationalize this by the lack of differentiation for transverse vectors
(see Fig.~\ref{fig:mol-vec}b). This has led us to the development of
CVs that are tailored to sPS.

Fig.~\ref{fig:mol-vec}b shows a longitudinal view of a chain
conformation, as well as transverse sections of the two main
polymorphs of interest: the experimentally-determined $\alpha$ and
$\beta$ structures~\cite{Guerra1990}. We herein propose two sets of
vectors: one longitudinal vector, ${\bm v}_1$, and one transverse
vector, ${\bm v}_2$. ${\bm v}_1$ is oriented along the backbone, it is
defined by the interparticle vector between two consecutive monomers
(i.e., backbone beads). As for the transverse vector ${\bm v}_2$, for
each monomer, it originates from the second backbone and point to the
bisector of the two closest side chains. Fig.~\ref{fig:mol-vec}b
indicates that typical angles for the $\alpha$ and $\beta$ polymorphs
are roughly 120$^\circ$ and 180$^\circ$, respectively. Note that here,
to improve the efficiency of computation, the side-chain vector is an
average over each whole molecule. Effectively this assumes a
homogeneous configuration across the chain, in line with the system
sizes considered here.

Because these two polymorphs consist of identical chain conformations,
no differences can be extracted from \emph{intrachain} statistics.
Instead, differences between polymorphs are concentrated in the
\emph{interchain} configurations. As such, any CV that is to
distinguish between polymorphs ought to focus on interchain
geometries. By symmetry, the most noticeable differences occur along
the transverse sections. These differences can be displayed more 
intuitively by the distribution of angles between neighboring 
transverse vectors ${\bm v}_2$ (see Fig. S1), which is from unbiased 
MD simulations at 300~K. 
In the following, we will introduce two kinds of CVs which are
functions of these transverse angles. 

\subsubsection{Legendre polynomial $P_2$} 

The second Legendre polynomial $P_2$ probes the orientation between
two vectors ${\bm e}_i$ and ${\bm e}_j$
\begin{equation} \label{eq:p2}
    P_2({\bm e})=\dfrac{3}{2}({\bm e}_i\cdot {\bm e}_j)^2-\dfrac{1}{2}.
\end{equation}
$P_2$ is a natural candidate to describe orientational ordering and
has been used to monitor the crystalline growth and/or state of
polymer chains~\cite{Ko2004, Waheed2005}. Some of us previously used
$P_2$ to monitor the melting transition of sPS~\cite{Liu2018}. 

\subsubsection{SMAC} 
    
Giberti et al.~\cite{Giberti2015} recently introduced a CV to capture
the inherent variety of crystal symmetries~\cite{Santiso2011}. This
CV, simply called SMAC in {\sc Plumed}, is formulated with the aim of
accounting for both local density and the mutual orientation of
molecules. In our case, it probes features of the angle distributions
between transverse vectors, as shown in Fig. S1. It compares an input
angle, $\theta_{ij}$, to a reference angle, $\theta_n$, via
\begin{equation} \label{eq:kernel}
    K_n (\theta_{ij}-\theta_n)=e^{-\left((\theta_{ij}-\theta_n)^2/2\sigma_n^2\right)}.
\end{equation}
This kernel smoothly interpolates from identical to distant angles
leading to values from 1 to 0, respectively.  SMAC relies on these
kernels to probe one or multiple reference angles according to the
phase of interest, supplemented by a smooth cutoff scheme
\begin{equation} \label{eq:smac}
    s_i= \frac{\left\{1-\Uppsi\left[\sum_{j\ne i}\sigma(r_{ij})\right]\right\}
    \sum_{j\ne i}\sigma(r_{ij}) \sum_n K_n (\theta_{ij}-\theta_n)}
    {\sum_{j\ne i}\sigma(r_{ij})}.
\end{equation} 
Eq.~\ref{eq:smac} relies on the two switching functions $\sigma(r) =
\frac{1 - \left(r/r_\sigma\right)^6 }{ 1 -
\left(r/r_\sigma\right)^{12}}$ and $\Uppsi(r) = \exp(-r/r_\Uppsi)$,
where $r_\sigma$ and $r_\Uppsi$ (see Tab.~S1)
provide a balance to keep the CV local, while incorporating enough
numbers. The quantity is then averaged over all $s_i$:
$\mathcal{S} = \sum_{i=1}^{N} s_i / N$.

Based on the angle distributions of Fig. S1, we construct a number of
SMAC CVs using various sets of reference angles to optimally
distinguish between sPS phases. Tab.~S1
lists the set of CVs we used in this work. For instance, the main
peaks for the $\alpha$ and $\beta$ phases led to the definition of
SMAC CVs $\mathcal{S}_\alpha$ and $\mathcal{S}_\beta$, centered at 120
and 170$^\circ$, respectively. To emphasize both features at once, we
constructed a CV based on their difference: $\Delta \mathcal{S} =
\mathcal{S}_\beta - \mathcal{S}_\alpha$. This can be observed through
a monitoring of the CV during unbiased MD simulations (Fig.~S2c). 
Three additional SMAC CVs are presented in Tab.~S1.

\subsection{Metadynamics}
\label{sec:metad}

Well-tempered metadynamics simulations were performed at
400\,K~\cite{Barducci2008, Laio2002, Barducci2011, Abrams2014,
Dama2014, Tiwary2014, valsson2016enhancing}. The bias deposition
stride was set to be 0.5 ps, and the bias factor was 20. The
Gaussian-bias width was set to 0.01 for $\Delta\mathcal{S}$, and 0.05
otherwise. The initial Gaussian heights were all set to 3\,kJ/mol.
Each simulation consisted of multiple walkers~\cite{Raiteri2006}: In
metadynamics, we refer to a biased run that explores the CV-space as a
walker, while multiple walkers simultaneously explore and carve the
same free-energy landscape. This method can significantly speed up the
convergence of the simulations, as all walkers contribute to a single,
combined free-energy landscape. In this work, 2 walkers were run in
parallel and initialized from the $\alpha$ and $\beta$ forms,
respectively. All simulations were performed using {\sc Gromacs}
5.1.4~\cite{Abraham2015} and {\sc Plumed} 2.4~\cite{Tribello2014,
plumed-nest}. Simulations were carried out in the isothermal-isobaric
ensemble at $P=1$~bar using the velocity-rescale
thermostat~\cite{Bussi2007} and the anisotropic Parrinello-Rahman
barostat~\cite{Parrinello1981}. We ensured stable variations in the
simulation box by restraining the range of allowed geometries
($5<a^2<9;~5<b^2<9;~6<c^2<8$~nm$^2$). We also restrained the director
of each chain to lie within 30$\degree$ of the box's $z$ component.
This avoided significant collective rotations of the chains with
respect to the simulation box, leading to alignments along the other
coordinates. Such a scheme merely aims at enforcing all chains to
loosely lie within an arbitrarily-chosen $z$ axis. This helped avoid
artifacts when calculating longitudinal and transverse vectors,
especially at higher temperatures. More details can be found in the
SI.

\subsection{Quantum-chemical methods}
\label{sec:dft}
Quantum-chemical methods are used to simulate the relative
thermodynamical stability of the two sPS polymorphs. The two
polymorphic forms identified by the metadynamics simulations and
experimentally characterized in the literature have been taken as
starting points for local geometry optimization. Phonon modes are
computed to confirm the stationary points as local minima and to give
access to the temperature-dependent harmonic Gibbs free energy
\begin{align}
 G^{\text{HA}}(T,P)= E_{\text{latt}}+G_{\text{vib}}^{\text{HA}}(T)+PV\,.
 \label{eq:freeenergy}
\end{align}
Here, $E_{\text{latt}}(V)$ is the zero-temperature internal energy of
the crystal given per monomer unit---the lattice energy. The
vibrational contributions are
\begin{align}
 G_{\text{vib}}^{\text{HA}}(T)= 
 \sum_{\mathbf{k},p} \frac{\hbar\omega_{\mathbf{k},p}}{2}
 + k_{\text{B}}T\sum_{\mathbf{k},p}
 \left[\ln \left(1- \mathrm{e}^{
     -\frac{\hbar\omega_{\mathbf{k},p}}{k_{\text{B}}T}} \right)\right]\,,
\end{align}
where the phonon frequencies $\omega_{\mathbf{k},p}$ correspond to a
$\mathbf{k}$-point in first Brillouin zone and a phonon band index
$p$. The temperature-dependent harmonic enthalpy, $H^{\text{HA}}$, is
described in the SI. The quantum chemical calculations are performed
using density functional theory (DFT), which is the method of choice
for many materials applications due to its favorable accuracy to
computational cost ratio~\cite{Kieron-JCP, Becke-JCP, Truhlar-JCP,
dft-materials-rev}. The DFT calculations are done with a screened
exchange hybrid density functional, dubbed HSE-3c~\cite{hse3c,
3c_review}. It combines accurate descriptions of geometries over a
broad class of systems with an efficient treatment of non-local
exchange and long-range London dispersion
interaction~\cite{long_xbond, short_ch}. For a general overview of
dispersion corrections in the density functional framework and the
treatment of molecular crystals see Refs.~\onlinecite{rev-mf-disp,
chemrev_beran, vdw_perspective}. The implementation of HSE-3c into the
{\sc CRYSTAL17} program enables the fast computation of electronic
structures and phonon modes using all point- and space-group
symmetries~\cite{crystal17, crystal17_wire}. Geometry optimizations
are performed with tight convergence thresholds and in space groups
$P3_1$ ($\alpha$) and $Pnma$ ($\beta$). The Brillouin zone has been
sampled with a 1$\times$1$\times$5 and 1$\times$5$\times$3 grid for
the $\alpha$ and $\beta$ polymorphs, respectively. $\Gamma$-point
frequencies have been computed by sHF-3c~\cite{hf3c, s-hf3c}, above
$\Gamma$-point frequencies are tested to be negligible at the DFTB3-D3
level \cite{dftb, dftb2, dftb3d3}, which has been shown to be a
reliable approach for organic solids~\cite{qha_cbz}.


\section{Acknowledgment}
We thank Christine Peter for insightful discussions, as well as
Bingqing Cheng, Robinson Cortes-Huerto, and Hsiao-Ping Hsu for
critical reading of the manuscript. CL was supported by the Max Planck
Graduate Center. TB was partially supported by the Emmy Noether
program of the Deutsche Forchungsgemeinschaft (DFG).


\bibliographystyle{unsrt}

\bibliography{paper}


\end{document}